\documentclass[journal=ancac3,manuscript=article]{achemso}

\usepackage{amsmath,amssymb,amsfonts}
\usepackage{bm}
\usepackage{color, soul}
\usepackage{graphicx}
\usepackage{float}
\usepackage{wrapfig}
\usepackage{verbatim}
\usepackage[english]{babel}
\usepackage{hyperref}[colorlinks=true,bookmarks=false,citecolor=blue,urlcolor=blue]
\usepackage{natbib}
\usepackage{subfigure}
\usepackage{amstext, mathrsfs, textcomp}
\usepackage{multirow}
\usepackage{xcolor}
\usepackage{enumerate}
\usepackage{epstopdf}
\usepackage{tabularx}
\makeatletter

\def\dfrac{\displaystyle\frac} 

\usepackage{gensymb}

\setkeys{acs}{etalmode=truncate,maxauthors=0}
\graphicspath{{figs/}}
\author{Alexander Barulin$^*$}
\affiliation{Russian Quantum Center, Skolkovo, 143025 Moscow, Russia}
\author{Olesia Pashina$^*$}
\affiliation{School of Physics and Engineering, ITMO University, Saint-Petersburg 197101, Russia}
\author{Daniil Riabov}
\affiliation{School of Physics and Engineering, ITMO University, Saint-Petersburg 197101, Russia}
\author{Olga Sergaeva}
\affiliation{School of Physics and Engineering, ITMO University, Saint-Petersburg 197101, Russia}
\alsoaffiliation{Department of Information Engineering, University of Brescia, Brescia, 25123, Italy}
\author{Zarina Sadrieva}
\affiliation{School of Physics and Engineering, ITMO University, Saint-Petersburg 197101, Russia}
\author{Alexey Shcherbakov}
\affiliation{School of Physics and Engineering, ITMO University, Saint-Petersburg 197101, Russia}
\author{Viktoriia Rutckaia}
\affiliation{Martin-Luther-University of Halle-Wittenberg, 06120 Halle (Saale), Germany }
\alsoaffiliation{Advanced Science Research Center, City University of New York, 10031 New York, USA}
\author{J{\"o}rg Schilling}
\affiliation{Martin-Luther-University of Halle-Wittenberg, 06120 Halle (Saale), Germany }
\author{Andrey Bogdanov}
\affiliation{School of Physics and Engineering, ITMO University, Saint-Petersburg 197101, Russia}
\alsoaffiliation{Qingdao Innovation and Development Center of Harbin Engineering University, Qingdao 266404, China}
\author{Ivan Sinev}
\affiliation{School of Physics and Engineering, ITMO University, Saint-Petersburg 197101, Russia}
\alsoaffiliation{Experimentelle Physik 2, Technische Universit{\"a}t Dortmund, 44227 Dortmund, Germany}
\author{Alexander Chernov}
\affiliation{Russian Quantum Center, Skolkovo, 143025 Moscow, Russia}
\alsoaffiliation{Center for Photonics and 2D Materials, Moscow Institute of Physics and Technology, Dolgoprudny 141700, Russia}
\author{Mihail Petrov}
\affiliation{School of Physics and Engineering, ITMO University, Saint-Petersburg 197101, Russia}
\email{m.petrov@metalab.ifmo.ru}

\title{
Thermo-optical bistability enabled by bound states in the continuum in silicon  metasurfaces\\ 
}

\begin{document}
\def\thefootnote{*}\footnotetext{These authors contributed equally to this work}
\begin{tocentry}
\begin{centering}
\includegraphics[width=8cm]{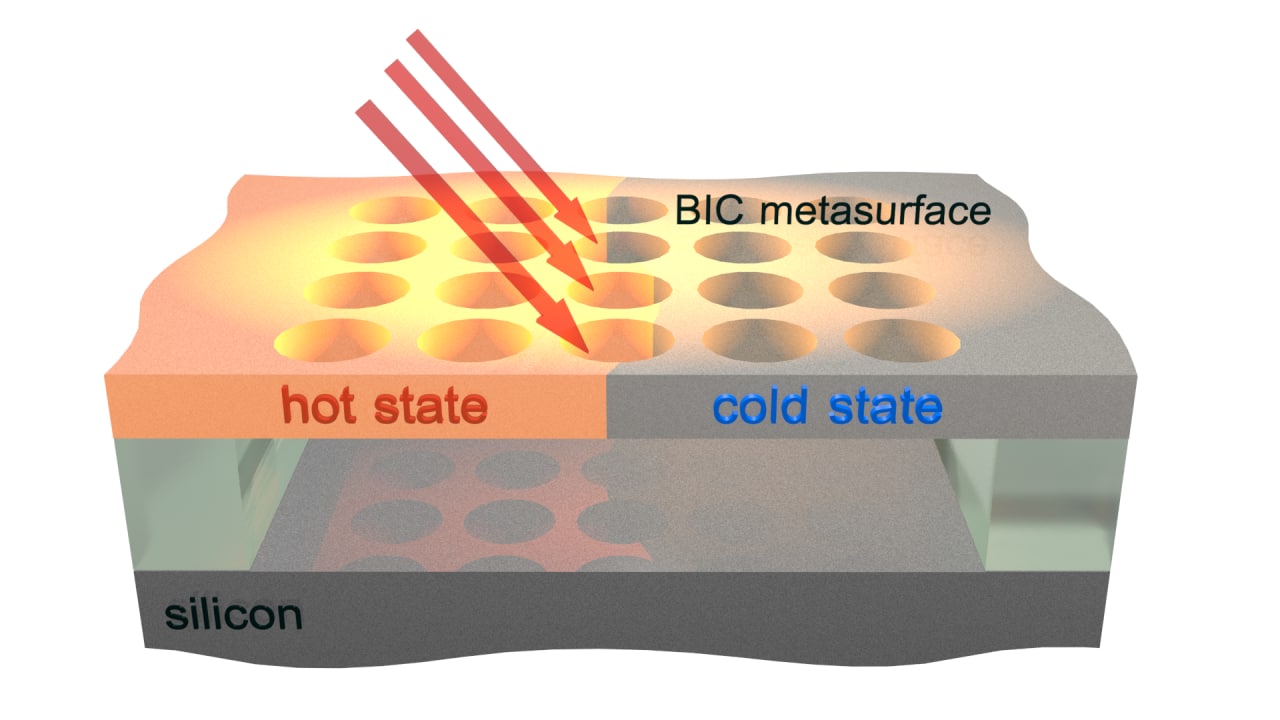}
\end{centering}
\end{tocentry}

\begin{abstract}

The control of light through all-optical means is a fundamental challenge in nanophotonics and a key effect in optical switching and logic.  The optical bistability effect enables this control and can be observed in various planar photonic systems such as microdisk and photonic crystal cavities and waveguides. However, the recent advancements in flat optics with wavelength-thin optical elements require nonlinear elements based on metastructures and metasurfaces. The performance of these systems can be enhanced with high-Q bound states in the continuum (BIC), which leads to intense harmonic generation, improved light-matter coupling, and pushes forward sensing limits. 
In this study, we report on the enhanced thermo-optical nonlinearity and the observation of optical bistability in an all-dielectric metasurface membrane with BICs. Unlike many other nanophotonic platforms, metasurfaces allow for fine control of the quality factor of the BIC resonance by managing the radiative losses. This provides an opportunity to control the parameters of the observed hysteresis loop and even switch from bistability to optical discrimination by varying the angle of incidence. Additionally, we propose a mechanism of {\it nonlinear critical coupling} that establishes the conditions for maximal hysteresis width and minimal switching power, which has not been reported before. Our work suggests that all-dielectric metasurfaces supporting BICs can serve as a flat-optics platform for optical switching and modulation based on strong thermo-optical nonlinearity. 


\end{abstract}

\section{Introduction}


Optically nonlinear elements provide unique functionalities of photonic devices allowing for all-optical modulation of light \cite{reed_silicon_2010,chen_all-optical_2020}, shaping of light pulses \cite{Divitt2019}, and even realisation of optical logic elements for classical \cite{amo_excitonpolariton_2010,ballarini_all-optical_2013} and neuromorphic \cite{shastri_photonics_2021,cheng_-chip_2017,mirek_neuromorphic_2021} optical computing and  simulations. Optical  bistability \cite{gibbs_optical_1987,Lugiato1983}, i.e. the ability of a system to be in two different states at the same excitation conditions, is one of the key features that enable the creation of optical logic devices such as optical transistors, switchers and  memory elements. Following the constant demand for miniaturisation of such devices, bistability was sought for and observed in various systems such as microring  resonators\cite{Jiang2020},  photonic crystal cavities \cite{gu_regenerative_2012} and waveguides \cite{chen_all-optical_2011}, nanobeam waveguides \cite{quan_ultrasensitive_2011} and others, compatible with the concept of integrated nanophotonics. In parallel, one could witness the rapid development of the so-called flat optics\cite{Capasso2018}, where classical optical elements are gradually substituted by wavelength-thin  metasurfaces and metagratings. Their tailored  optical properties and rich functionalities reproduce or even surpass those of bulk optics elements with the help of advances in nanostructuring technologies and novel materials\cite{tonkaev_multifunctional_2022}. The subwavelength thickness of metasurfaces require enhancement of light-matter interaction through utilizing high-$Q$ resonant states for observing nonlinear effects. On this path semiconductor optical metasurfaces provide great opportunities for engineered high-$Q$ resonant states with bound states in continuum (BICs), which are nonradiative modes embedded in the radiation continuum. From practical perspective, BIC metasurfaces have extremely large quality factor limited by nonradiative losses~\cite{PhysRevB.105.165425}, diffraction into a substrate~\cite{sadrieva2017transition}, sample footprint size and imperfections, such as disorder~\cite{Ni:17,maslova2021bound} and surface roughness~\cite{sadrieva2017transition}. BICs have recently shown their efficiency in boosting the performance of photonic systems \cite{koshelev_bound_2021} and achieving remarkable results in the enhancement of higher harmonic generation \cite{koshelev2020subwavelength} and nonlinear optical response \cite{sinev2021observation}.

In this paper, we employ high-$Q$ BICs for optical heating of  silicon metasurface membrane and enhancing thermo-optical nonlinearity finally reaching the bistability regime. While thermo-optical effects in semiconductor microcavities have been known for long time, they have recently attracted great interest due to the advancements of all-dielectric nanophotonics, which eventually led to the emergence of a standalone field of all-dielectric thermonanophotonics\cite{Zograf2021}. Controlled ohmic losses, large thermo-optical coefficients, and pronounced optical resonances in such structures enhance thermo-optical effects, leading to strong optical heating \cite{Zograf2017,Zograf2018}, decreasing of the onset of nonlinear regimes even in single subwavelength scatterers \cite{Duh2020,tang_mie-enhanced_2021}, and controlled phase transitions \cite{ruiz_de_galarreta_reconfigurable_2020,tonkaev_multifunctional_2022,Zograf2018}. 

We show that, since thermo-optical bistability regimes rely on a delicate balance between radiative and ohmic losses of the system, BIC metasurfaces offer a unique platform for controlling optical bistablity. In particular, we employ strong dependence of the optical mode linewidth on its wavevector in the vicinity of BIC to control the character of the nonlinear response from bitability regime  to optical discrimination regime. Additionally, one can modify the parameters of the hysteresis loop such as hysteresis width and threshold power by balancing the ratio between radiative and non-radiative losses, a unique attribute of BIC in metasurfaces. With that we put forward the {\it nonlinear critical coupling } condition for reaching maximal hysteresis width and minimal threshold power, and which is an alternative to the well-known classical critical coupling condition for linear systems.

\section{Results }

\begin{figure}[h!]
    \centering
    \includegraphics[width=0.99\columnwidth]{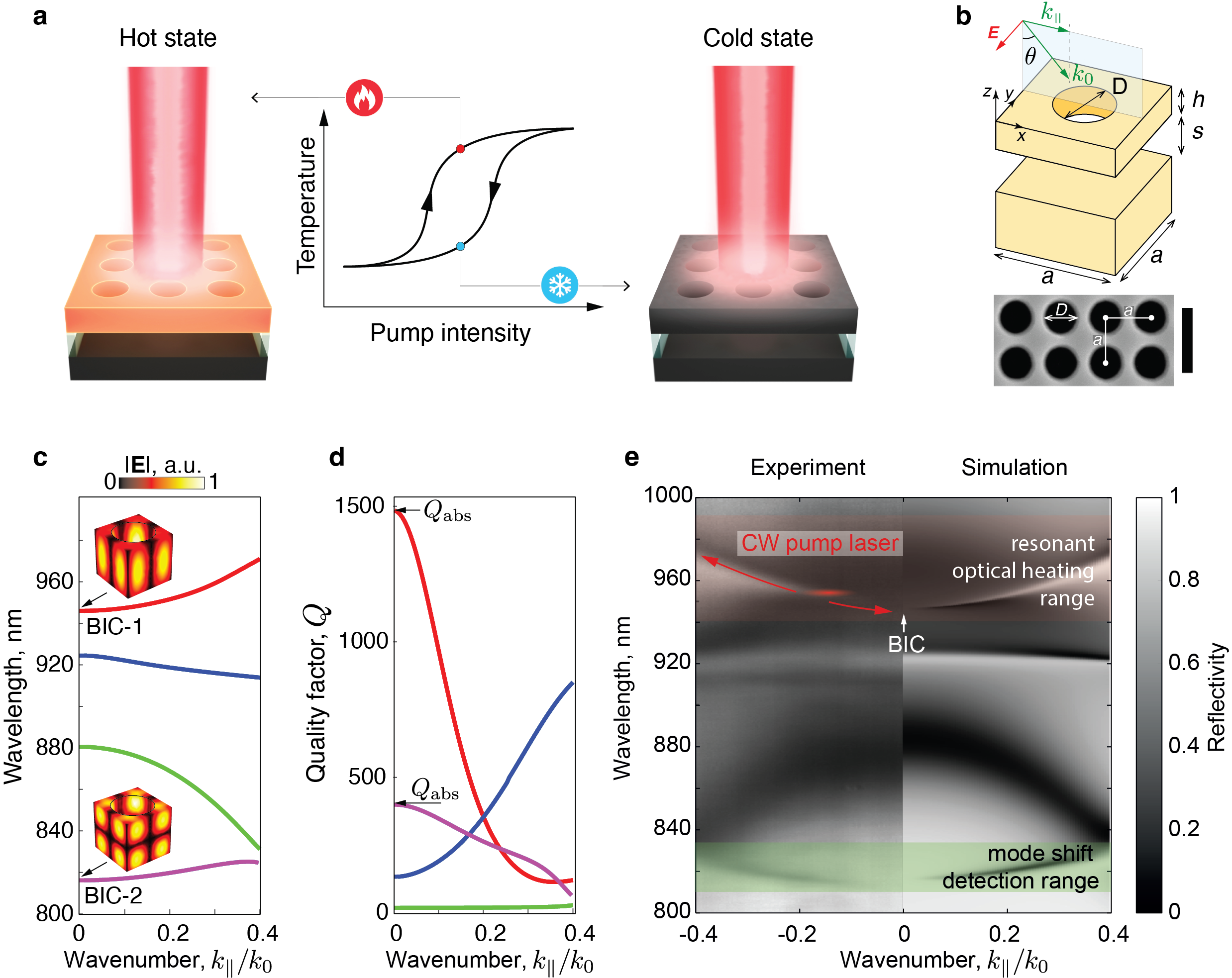}
    \caption{(a) Artistic view of the system: a suspended metasurface membrane with a high-$Q$ BIC optical mode  is excited with a CW laser beam at a small angle of incidence which results in its heating. Strong thermo-optical nonlinearity facilitated by the BIC resonance induces the bistability regime, when the metasurface can be either in ``hot" or ``cold" state for the same excitation conditions depending on the pre-history of the excitation. (b) The geometry of the system $D=310$ nm, $h=340$ nm, $s=2\ \mu$m, $a=355$ nm. The excitation is the TE polarization at the incident angle of $\theta$. (c-d) Angular dependence ($k_{||}/k_0$) of metasurface resonant wavelengths (c) and of corresponding quality factors $Q$ (d). BIC-1 and BIC-2 are two different BIC states excited in the metasurface for a given wavelength range. The quality factor $Q$ of BIC states at $\Gamma$-point is limited by absorption factor $Q_{\mathrm{abs}}$. The inset shows the distribution of the normalized electromagnetic field for BIC eigenmodes at $\Gamma$-point. (e)  Experimentally measured (left) and numerically calculated (right) angle-resolved reflectivity spectrum of the metasurface. BIC-1 is involved in the optical heating resides within 950-1000 nm range (resonant optical heating range, red). Temperature detection is performed through the measurements of the shift in the reflectivity dip of BIC-2 in the range 800-850 nm (mode shift detection range, green).}  
     \label{fig1:schematics}
\end{figure}

{The studied metasurface manifests itself in a rectangular hole array in silicon slab (photonic crystal slab)  with membrane configuration that blocks the heat dissipation channel through the substrate and further increases the optical heating efficiency [see Fig.~\ref{fig1:schematics} (a)].}  The metasurface was fabricated from silicon on insulator substrate with a 2~$\mu$m thick oxide layer using the multistage lithography and etching routine described in Methods. The final device is a $50\times50$ $\mu m^2$ membrane with slightly conical (8~degrees tilt angle) holes (See supplementary materials) with a top diameter of $D\approx$ 310~nm arranged in a square lattice with a period of $a=355$~nm [see Fig.~\ref{fig1:schematics} (b)]. The experiments were performed in the wavelength range 950--1000 nm where the structure is in metasurface     {(subdiffractive)} regime.

The fabricated membrane metasurface supports a symmetry-protected BIC-1 in the spectral range 950-1000 nm as can be seen from the numerical  simulation result shown in Fig.~\ref{fig1:schematics} (c,d) where the dispersion diagram and the $Q$-factor dependence over the first Brillouin zone are shown. The simulation results are in excellent agreement with  the angle-resolved reflectivity spectra shown in Fig.~\ref{fig1:schematics} (e): the spectral dip narrows when approaching the $\Gamma$-point until vanishing completely at normal incidence \cite{koshelev_bound_2021}. The experimental spectra are also in good agreement with reflectivity calculated with COMSOL Multiphysics package and  Fourier modal method \cite{Li1997, spiridonov_reformulated_2023} with account for the conical shape of the holes (also shown in Fig.~\ref{fig1:schematics}(e))
    
{\it Optical heating of the metasurface.} In the chosen spectral range, silicon has non-zero ohmic losses with complex refractive index $n(950
\ \mbox{nm})=3.591+1.18\cdot 10 ^{-3}$,~\cite{green_self-consistent_2008}  allowing for optical heating of the structure through the resonant excitation of BIC mode. For efficient excitation of the metasurface at a certain angle, we irradiated it with a large spot size ($\approx$ 20 $\mu$m FWHM) laser beam impinging the surface with low angular divergence. The angle of incidence was monitored with a spectrometer imaging the angle-resolved reflectivity map of the sample (laser spot in this configuration is visualized with a red oval in Fig.~\ref{fig1:schematics} (e)). Precise control of the laser wavelength (see also Methods) allowed us to measure the heating characteristics of the membrane for different initial detunings of the excitation with respect to the spectral position of the BIC mode for the chosen angle of incidence.

\begin{figure}[h!]
    \centering
    \includegraphics[width=0.85\columnwidth]{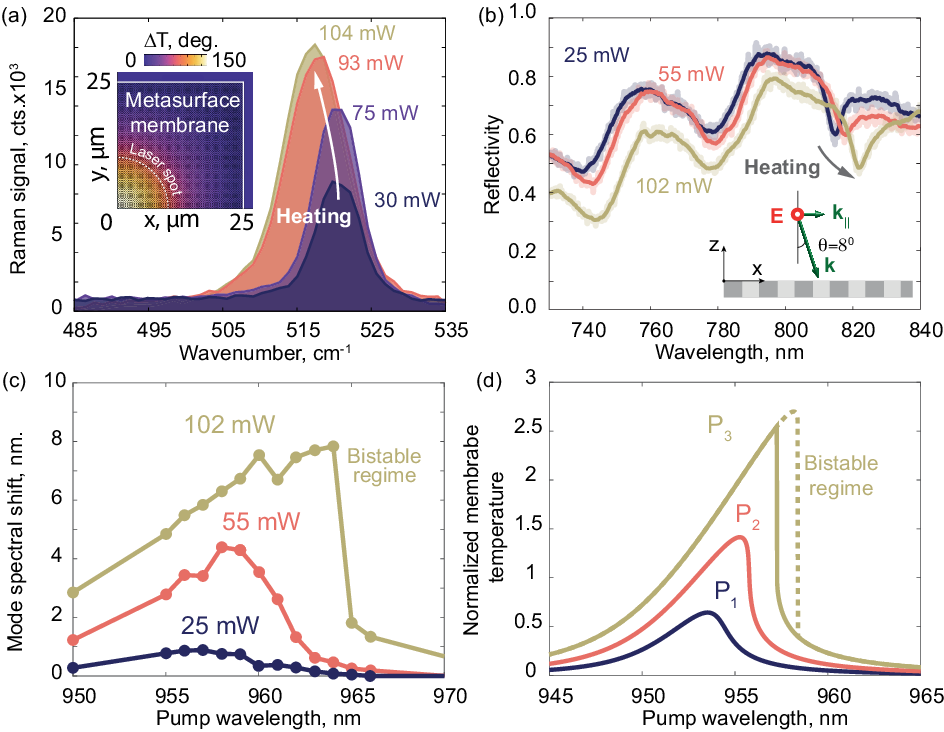}
    \caption{(a) The temperature-induced Raman line shift measured for a pump wavelength of 958~nm incident on the membrane at $\theta = 4$~degrees ($k_{||}/k_0$=0.07). Inset: calculated temperature distribution in the metasurface membrane under laser beam heating. (b) Evolution of the reflectivity spectra of the metasurface heated with different CW laser power at a wavelength of 960~nm and angle of incidence of $\theta = 8$~degrees ($k_{||}/k_0$=0.14).  (c) Experimentally measured spectral shift of the mode in the reflectivity spectra of the membrane near 815~nm for different pump wavelengths at 3 different pump powers. The excitation geometry is the same as in (b). Bistable regime is reached for a pump power of 102~mW. (d) Calculated spectral dependence of the temperature of the membrane on the wavelength of the pump laser for increased laser powers $P_1, P_2$, and $P_3$. At high power ($P_3$, red) the Lorentzian shape exhibits an abrupt drop indicating the onset of the bistability regime.}
     \label{fig2:heating}
\end{figure}

The suspended membrane configuration prevents  heat transfer in the direction normal to the surface resulting in  efficient  heating of the sample. The simulated temperature distribution under the optical excitation is shown in Fig.~\ref{fig2:heating} (a). The  heating efficiency in the center of the laser spot was as high as 5-8.5 K/mW as estimated from the modelling in COMSOL Multiphysics package with the optical heating module. 
 
\begin{figure}[h!]
    \centering
    \includegraphics[width=0.99\columnwidth]{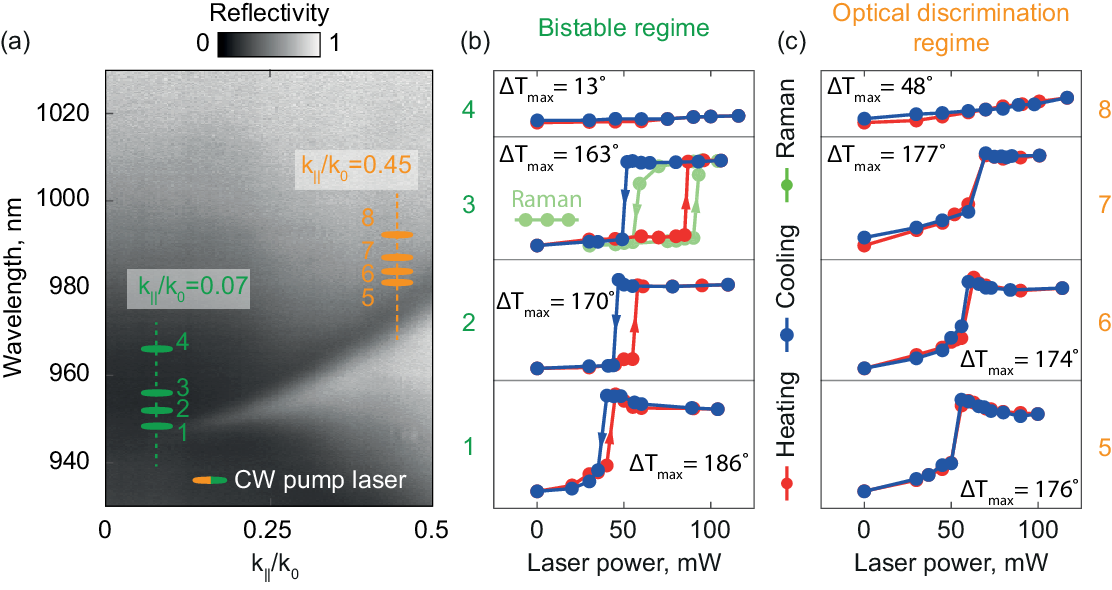}
    \caption{(a) Reflectivity map in $\omega$-$k$ space with excitation parameters marked for hysteresis regime (points 1-4, green) and for optical discrimination regime (points 5-8, orange).  (b,c) Evolution of the membrane temperature as a function of the laser power  for increasing (red) and decreasing (blue) sequence: (b) hysteresis regime, points 1-4,  and (c) discrimination regime, points 5-8. The temperature readings were extracted from reflectivity spectrum shift. The green curve for point 3 corresponds to the temperature obtained from Raman thermometry measurements.  }
     \label{fig3:bistability}
\end{figure}

Experimentally, the temperature increase induced by the laser was determined via two approaches. The first involved monitoring the red spectral shift of another optical mode of the membrane (BIC-2) at $\sim $815~nm (see Fig.~\ref{fig1:schematics} (e) for the specified temperature measurement region, and Fig.~\ref{fig2:heating} (b) for measured data) induced by the temperature-dependent change of the refractive index of silicon $n(\Delta T)=n+n_2\Delta T$. This mode is also a BIC, thus, the reflection was measured at non-zero angle $k_{||}/k_0=0.14$ ($\theta=8^{\circ}$). In the spectral range of temperature control (Fig.~\ref{fig1:schematics} (e), green area), we approximated the thermorefractive coefficient of silicon as  $n_2=3.315\cdot10^{-4}~\mathrm{K^{-1}}$~\cite{Jellison1986Jul}. In the second approach, concurrently with the optical heating routine, we measured the Raman scattering signal and extracted the temperature change from the spectral shift and broadening of the Raman peak according to the well-known dependence\cite{Balkanski1983}.  Raman thermometry was used to verify the obtained temperature measurements as shown in Fig.~\ref{fig3:bistability} (b), however, the main results were obtained with the first method as it is much more convenient in the utilized experimental geometry.   

{\it Optical bistability.}  The first indication of the thermo-optical bistability in the metasurface is revealed in the measured dependencies of the structure temperature on the laser wavelength as it is swept across the resonant wavelength of the BIC-1 mode. Such curves measured at different laser powers are shown in Fig.~\ref{fig2:heating} (c)  for excitation at $k_{||}=0.14 k_0$. At low laser power, the spectral shift of the mode directly related to temperature change through the thermorefractive effect is weak and decreases symmetrically for both negative and positive detunings from the resonant wavelength of the mode - in this case, 956~nm. However, as the laser power increases, the maximum temperature grows and the initial Lorentzian shape of the temperature dependence becomes highly asymmetric until at $\approx$ 102 mW it exhibits an abrupt drop of temperature at 965~nm which is indicative of the onset of bistability regime. Note here that since the laser output has to be closed during the wavelength tuning process and the membrane cools down between the measurements, the mode spectral shift, in this case, does not fully reproduce the theoretical bistability curve shown in Fig.~\ref{fig2:heating} (d) and obtained within the single mode stationary model \cite{ryabov_nonlinear_2022} as described below in Discussion section.  The appearance of a hysteresis loop in the simulations is a direct evidence of the bistability regime.

We further perform a more direct experimental demonstration of the bistability by cycling the laser power in increasing and then in decreasing sequence at certain initial spectral detunings between the laser line and the optical mode. Excitation conditions (excitation wavelength and incident angle) for these measurements are illustrated Fig.~\ref{fig3:bistability} (a) on top of the angle-resolved reflectivity map measured in the linear regime. Points 1-4 correspond to a set of measurements taken for $k_{||}=0.07k_0$ and different  values of detuning between the laser frequency and the BIC mode frequency. As expected from the simulation, the membrane temperature evolution upon the increase and consequent decrease of the intensity exhibits a hysteresis loop with height  and width dependent on the initial detuning as shown in Fig.~\ref{fig3:bistability} (b). Here, for  point 3 we also show the comparison of data extracted from Raman thermometry with the reflection thermometry which confirms good correspondence between the two methods (slight difference in the hysteresis bounds is due to minor variations of the excitation conditions). 

 The heating curves demonstrate strong dependence not only on the detuning, but also on the angle of incidence. The latter, as we show further, is facilitated by strong dependence of radiative losses of the membrane optical mode on the $k$-vector due to the presence of BIC. As an illustration to that,
we perform the same experiments for different angle of incidence ($k_{||}=0.4k_0$), points 5-8 in Fig.~\ref{fig3:bistability} (a). Here, for the same values of detuning as for points 1-4, the system supports a very distinct nonlinear regime. The hysteresis loop collapses (Fig.~\ref{fig3:bistability} (c)), and the temperature dependence demonstrates a sharp discontinuity at a critical value of incident power. Such behaviour corresponds to optical discrimination regime, when either low or high power signal can be filtered out in reflectance or transmittance~\cite{Hirano1999May, Brzozowski2001Jan}.

\section{Discussion}

\begin{figure}[h!]
    \centering
    \includegraphics[width=0.99\columnwidth]{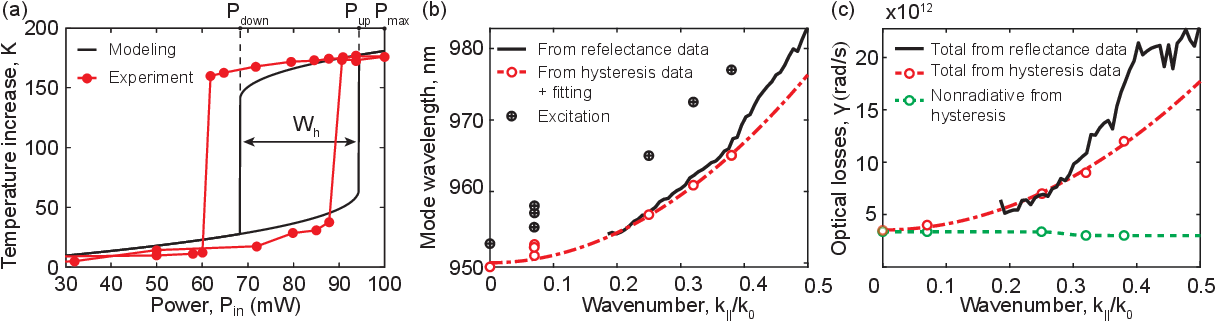}
    \caption{(a) Dependence of metasurface temperature increase on the excitation intensity in the bistability regime. The typical parameters of the hysteresis loop are marked: hysteresis width ($W_{\text{h}}$), maximal available excitation intensity ($P_{\text{max}}$),  upward/downward transition power ($P_{\text{up/down}}$). Results provided by the theoretical model are shown by the black solid line.  The excitation parameters are $k_{||}/k_0 = 0.32$, $\lambda_0 = 960.8$ nm, $\lambda = 972.5$ nm, $\Delta\omega = 2.36\cdot 10^{13}$ rad/s. The extracted parameters of non-radiative and radiative losses rates are $\gamma_{nr} = 3.37\cdot 10^{12}$ rad/s, $\gamma_{r} = 5.97\cdot 10^{12}$ rad/s. (b) Spectral position of the BIC mode extracted from the reflectivity measurements (black solid line), and from  fitting the hysteresis data at different $\Delta \lambda$ and $k_{||}$ (hollow circles and dashed fitting curve). The crossed circles denote the pump wavelengths used in the experiment. (c) The total optical losses of BIC mode extracted from the reflectivity spectra (solid black line). The total (red) and non-radiative (green) losses obtained from fitting the hysteresis loops (hollow circles) and their interpolation curves (dashed lines).}
     \label{fig4:Fittings}
\end{figure}
As we emphasized above, the nonlinear response of the heated metasurface strongly depends on the excitation angle due to $k$-dependent radiative losses and coupling strength to the BIC mode. At this point, we devise a strategy for reaching particular parameters of nonlinear response. In the bistability regime, those parameters are the switching powers from cold to hot states ($P_{\text{up}}$) and backwards ($P_{\text{down}}$) (see Fig.~\ref{fig4:Fittings} (a)) and hysteresis width $W_h= P_{\text{up}}-P_{\text{down}}$. Engineering of {thermo-optical switching} implies reaching certain values of these parameters and tradeoffs between them. 

Our further analysis based on a single mode nonlinear model allows for  identifying the hysterersis parameters $W_{h}$  and $P_{\text{up}}$. Introducing the amplitude of the mode $a$, one can then utilize the coupled mode theory equation for harmonic excitation~\cite{haus1984waves}

\begin{equation} 
    a =\dfrac{i\sqrt{\gamma_r} {f}}{i\Delta \omega+\gamma} 
    \label{eq:amp_init}
\end{equation}

where $\Delta \omega=\omega_0-\omega$ is the detuning frequency between the mode frequency  $\omega_0$ and laser frequency $\omega$,  $\gamma = \gamma_{r}  + \gamma_{nr}$ is the total loss rate which is given by radiative ($\gamma_r$) and nonradiative ($\gamma_{nr}$) channels of dissipation, and  ${f}$ corresponds to the amplitude of the incident wave ($|f|^2$ is the power of the incident wave). The nonlinearity of the system then can be taken into account by substituting the intensity dependent mode frequency  $ \omega_0 \rightarrow \omega_0 - \alpha|a|^2$, where $\alpha>0$ defines  the optical nonlinearity  of the metasurface.\cite{ryabov_nonlinear_2022}  The equation describing the energy stored in the cavity then takes the form   

\begin{align}
 \label{eq:mode_ampl_nonlin}
     |a|^2 = \frac{\gamma_r|f|^2}{\gamma^2 + \left(\Delta\omega-\alpha |a|^2\right)^2},
\end{align} 
 which is a third-order algebraic equation responsible for the appearance of multiple steady-states in the system.  Its analysis (see Supplementary Materials) provides one with the explicit expression for the hysteresis loop width $W_h$ and switching power $P_{\text{up}}$ for a given $\gamma_r$,$\gamma_{nr}$, $\Delta \omega $, and $\alpha$:

\begin{equation} \label{eq:hyst_width}
\begin{aligned}
    W_{\text{h}} = \frac{4}{27} \dfrac{\gamma^3}{\alpha \gamma_r}\left(\dfrac{\Delta{\omega}^2}{\gamma^2} - 3\right)^{3/2},\qquad P_{\text{up}} = \frac{2}{27}\dfrac{\gamma^3}{\alpha \gamma_r}\left( \dfrac{\Delta{\omega}}{\gamma}\left[\dfrac{\Delta{\omega}^2}{\gamma^2} + 9\right]\right) + \dfrac{W_h}2.
\end{aligned}
\end{equation} 

We applied this model to describe the experimentally observed bistable BIC-driven optical heating. For that, we found the model parameters $\Delta \omega$, $\gamma$, and $\gamma_r$ by fitting the experimentally measured hysteresis parameters $P_{\text{up}}$ and $W_h$. An example of such fitting is shown in Fig.~\ref{fig4:Fittings} (a), where we compare the measured hysteresis loop and the theoretical curve calculated for the experimental conditions. This procedure repeated for various $k$-vectors allowed us for obtaining the dispersion line $\omega(k)$ of BIC at the particular angles where the  hysteresis was observed (shown in Fig.~\ref{fig4:Fittings} (b) with red circles), and for total and radiative losses shown  shown in Fig.~\ref{fig4:Fittings} (c) with red and green circles correspondingly. One may note, that the values of non-radiative losses $\gamma_{nr}$ of BIC are almost constant over the Brillouin zone, which agrees well with the results of the modelling. We have verified the  parameters of detuning and losses found from the hysteresis analysis with the data directly obtained from 
 the angular dependent reflectivity map shown in Fig~\ref{fig1:schematics} (b) using the Fano lineshape fitting. The extracted $k$-dependence of the mode wavelength $\lambda_0=2\pi c/\omega_0$ is shown in Fig.~\ref{fig4:Fittings}~(b) with a black solid line. Close to the $\Gamma$-point, the reflectivity modulation vanishes due to the low radiative losses natural for BIC, which hinders the fitting procedure. Instead, we extrapolated the $\omega_0(k)$ curve with quadratic dependence. The fitting procedure, however, does not allow to separate radiative and non-radiative losses and only total losses can be found from the reflectivity spectra fitting. One can see that the data obtained from the reflectivity map and from the hysteresis analysis agrees well with each other.  
\begin{figure}[h!]
    \centering
    \includegraphics[width=0.99\columnwidth]{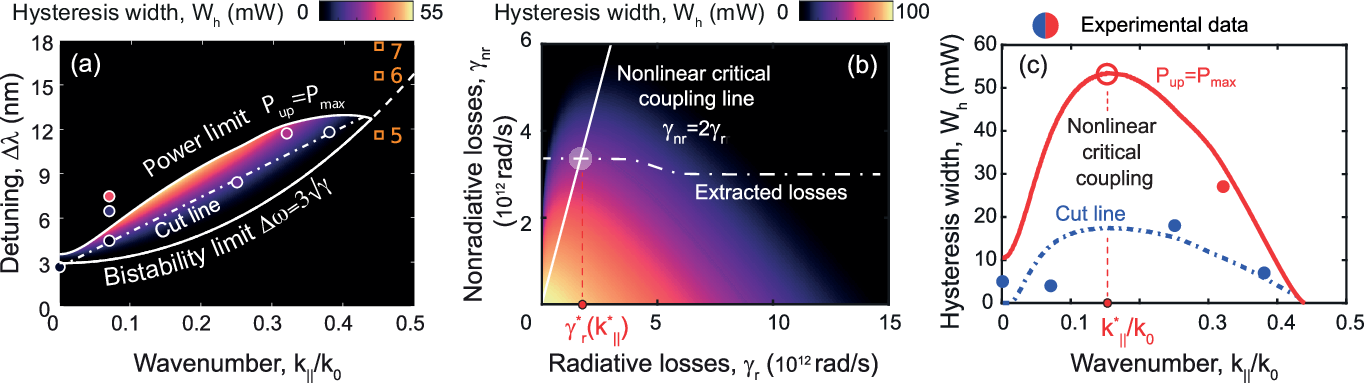}
    \caption{ (a)  
    Calculated map of the hysteresis loop width for different detunings and angles of incidence. Solid line encircles the $\Delta \lambda, k$ region where the bistability regime is expected to be observed in the experiment. The minimal detuning line $\Delta \omega=\sqrt{3}\gamma$ is extended with a dashed line. An additional cut line (dash-dot) tracks the available experimental points shown with circles. The color of each circle shows the measured hysteresis width. (b) Map of the hysteresis width as a function of radiative and non-radiative losses  manifesting the nonlinear critical  coupling regime: the maximal width is observed for $\gamma_{nr}=2\gamma_r$  (c)Measured hysteresis widths (blue dots) and the hysteresis widths calculated for excitation conditions along the cut line (blue dased line). Red line shows the calculated loop width for maximal available pump power. The condition of nonlinear critical coupling is shown with a vertical line}
     \label{fig5:OptimalRegimes}
\end{figure}

{\it Nonlinear critical coupling condition.} 
The bistability regime manifests at certain conditions for spectral detuning and mode losses. These conditions can be specified in the   $\Delta \omega-k$  plane for BIC metasurface with the help of the model  provided by Eq.~\ref{eq:hyst_width}. The first constraint is the well-known condition for the minimal detuning frequency $\Delta \omega>\Delta \omega_{cr}=\sqrt{3}\gamma$ \cite{soljacic_optimal_2002,liu_analytical_2013, ryabov_nonlinear_2022} where $\Delta \omega_{cr}$ is the critical detuning value. We use the values of $\gamma (k)$ and $\lambda_0(k)$ (resonant mode wavelength) extracted from the experiments (Fig.~\ref{fig4:Fittings}) to show this condition in Fig.~\ref{fig5:OptimalRegimes}~(a) as a dependence of the respective wavelength detuning $\Delta \lambda=\lambda-\lambda_0(k)$ on the angle of incidence. In principle, every nonlinear resonator can be driven into the bistable regime if the detuning exceeds this critical value threshold. However, real systems exhibit another important constraint which is the limited excitation power. Indeed, from Eq.\ref{eq:hyst_width} one can see that the minimal power $P_{cr}$ for reaching bistability ($P_{\text{up}}(\Delta \omega_{cr})=P_{\text{down}}(\Delta \omega_{cr})$) is given by      

\begin{equation} 
P_{cr}=8\sqrt{3}\gamma^3/(9\alpha\gamma_r)
\label{eq:CriticalPower}
\end{equation} 

which indicates that the required power grows rapidly with losses rate $\gamma$. Furthermore, to observe the hysteresis loop, $P_{up}$ should also be within the capabilities of the laser. We plot this power restriction curve $P_{up}= P_{max}$ (which is $\approx$ 100 mW in our experiments) in Fig.~\ref{fig5:OptimalRegimes} (a) with another solid line. Together with the critical detuning curve they encircle the $\Delta \lambda- k$ range for which the bistability regime can be reached in the experiment. Overall, lowering of the critical power is desired for many applications since it obviously  decreases the energy consumption. However, as one can see from the Eq.~\ref{eq:CriticalPower}, minimizing the critical power requires simultaneous management of the radiative and non-radiative losses. For a given $\gamma_{nr}$, the minimal critical power $P_{cr}$ required for reaching the bistable state corresponds to the condition $\gamma_{r}=\gamma_{nr}/2$  as immediately follows from \eqref{eq:CriticalPower}. We refer to it as a {\it  nonlinear critical coupling} condition in full analogy to the well-known critical coupling condition in \textit{linear} systems $\gamma_{r}=\gamma_{nr}$ \cite{Zograf2021}.

Next, we turn to the analysis of the width of the hysteresis loop. Fig.~\ref{fig5:OptimalRegimes} (a) shows the map of the hysteresis width $W_h$ calculated based on the extracted data $\gamma(k),\lambda_0(k)$. The round scatter points each corresponds to the heating curve measured at certain values of $\Delta \lambda$ and $k_{||}/k_0$. The color of the scatter points corresponds to the experimentally measured width of the hysteresis. {One can note a divergence of the experimental data with the modelling at low angles of incidence. Two main factors contribute to this discrepancy. The first is the possible inaccuracy in determination of total losses and mode wavelength, since fitting of the mode lineshape in this range of $k_{||}$ is impossible due to low modulation depth in reflectivity: this leads to narrower loops observed in the experiment. The second one is the increase of the excitation power density for lower angles of incidence due to the better focusing and lower losses in the objective. This leads to hysteresis loops observed at higher detuning values. To avoid further complicating the picture, we did not account for this effect in the simulation, which was done for constant power density.} In Fig.~\ref{fig5:OptimalRegimes} (a) we also show points 5-7 that correspond to those in  Fig.~\ref{fig3:bistability} (a). Even though data points 6 and 7 lie above the bistability existance line (dashed line extension), the bistability is not observed for these parameters due to losses extraction uncertainty at the higher angles   (Fig.~\ref{fig4:Fittings}~(c)). Moreover, inhomogeneity of incoming power along the incident angle axes, as mentioned above, could also result into this discrepancy.

Fig.~\ref{fig5:OptimalRegimes} (a) further shows that the width of the hysteresis $W_h$ reaches its maximal value close to the power restriction curve (maximal detuning condition) and for a particular value of $k_{||}/k_0\approx 0.12$. This is  a clear evidence of nonlinear critical coupling. Indeed, one can derive the condition of maximal hysteresis width $W_h$  by setting $P_{up}=P_{max}$  and maximising expression Eq.~\ref{eq:hyst_width} for $W_h$. The derived conditions  again provide nonlinear critical coupling condition $\gamma_{r}=\gamma_{nr}/2$ (see  Supplementary Materials). Since  $\gamma_r$ for  BIC mode is varied along the Brillouin zone in wide range from 0 to $2\div3\ \gamma_{nr}$ while non-radiative losses stay almost constant (see Fig.~\ref{fig4:Fittings} (c) and Fig.~\ref{fig5:OptimalRegimes} (b)), the nonlinear critical coupling condition is satisfied at some particular value of $k_{||}/k_0\approx 0.12$. We have plotted the dependence of $W_h$ on radiative and non-radiative losses for detunings corresponding to $P_{up}=P_{max}$ in Fig.~\ref{fig5:OptimalRegimes}  (b) and labeled the intersection area correspondent to nonlinear  critical coupling condition. Two cross sections of the map Fig.~\ref{fig5:OptimalRegimes} (a) are shown in (c): one is along the power limitation line and another for the cut line that tracks the experimental data. Both exhibit a maximum at around $k_{||}/k_0\approx 0.12$ that corresponds to the nonlinear critical coupling. In the experiment, we reach the maximum hysteresis width at $k_{||}/k_0$=0.25. As one can see in Fig.~\ref{fig5:OptimalRegimes} (b,c), the measured width is ultimately very close to the optimal width for nonlinear critical coupling condition.

\section{Conclusion}

In this work, we have realised the thermo-optical bistability regime in a metasurface supporting high-Q bound states in the continuum. By varying the spectral detuning between the laser and  mode frequency and changing the angle of incidence excitation one can modify the optical response of the metasurface switching it from hysteretic response to optical discrimination regime. Moreover, varied radiative losses of quasi-bound states in the continuum allows to optimize the width of hysteresis loop, which corresponds to the regime of nonlinear critical coupling condition.  

\section{Methods}

\textit{Fabrication of the membrane} Membranes supporting bound states in the continuum were fabricated from silicon-on-insulator substrates with 2~$\mu m$ silicon oxide layer and 340~nm thick silicon slab on top. The silicon-on-insulator (SOI) substrate consisted of a 2 micrometer buried oxide layer covered with 340 nm Si layer. Arrays of holes were created using standard electron-beam lithography and inductively coupled plasma reactive ion etching (ICP RIE) techniques. The underlying oxide was etched with buffered hydrofluoric (HF) acid, resulting in a membrane symmetrically surrounded by air above and below.

\textit{Experimental measurements} Initial characterization of the optical mode structure of the membrane was performed with angle-resolved reflection spectroscopy. The membrane was excited through a high-NA objective (Mitutoyo Plan Apo NIR 100x, NA=0.7). The back focal plane of the objective was then imaged on the entrance slit of the spectrometer (Andor Kymera) which allowed to capture the reflectivity maps in ``wavelength - angle of incidence'' axes in a single shot.

Optical heating measurements were performed with a tunable continuous wave laser (H{\"u}bner Photonics C-Wave), which we used to excite the membrane at a controlled angle of incidence by focussing the linearly polarized laser beam at a particular point in the back focal plane of the objective.

\begin{acknowledgement}

We thank Yuri Kivshar for fruitful discussions. The work was supported by the Priority 2030 Federal Academic Leadership Program. I.S. acknowledges Mercur Foundation (Grant Pe-2019-0022) and TU Dortmund core funds.    

\end{acknowledgement}




\bibliography{Thermohysteresis.bib}

\end{document}


\section{Membrane geometry }
\begin{figure}[h!]
    \centering
    \includegraphics[width=0.99\columnwidth]{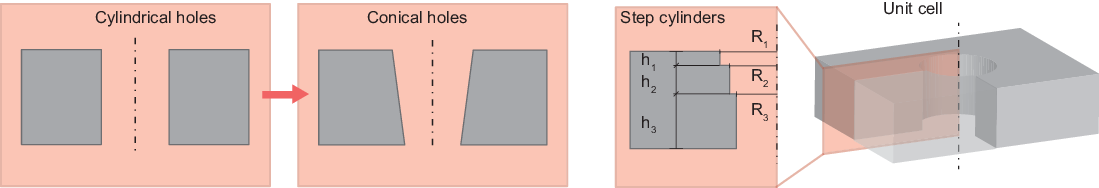}
    \caption{The geometry of the unit cell used in the modelling to obtain correspondence with the measured data in Fig.1 (e) in the main text. The parameters are close to the initially designed: $h_1=34$ nm, $h_2=102$ nm, $h_3=204$ nm, $R_1=157.5$ nm, $R_2=150$ nm, $R_3=137.5$ nm.}  
     \label{fig1_supp:cone}
\end{figure}

The membrane metasurface was fabricated using the e-beam lithography approach, which includes the plasma etching process. As a result, the shape and the size of the holes in the  array appear to differ from the initially designed. Thus, in order to match between the measured reflectivity spectra and modeling results, we had to take into account the conicity of  the holes appearing during the plasma etching process.  The optimized geometry is shown in Fig.~\ref{fig1_supp:cone} where the conical shape of the holes was modelled by three cylindrical steps of varied radius. Such geometry, on one hand, allows for taking into account the conicity of the structure and, on the other hand, is suitable for the Fourier modal method modelling.     

The obtained fitting parameters of  radiuses agree well with the designed diameter of the holes $D=310$ nm.

\section{Hysteresis parameters}
In order to define the hysteresis parameter, we start from Eq.~(2):
\begin{equation}
    \zeta(|a|^2) = |a|^2\cdot\left( \gamma^2 + \left(\Delta\omega-\alpha |a|^2\right)^2 \right) = \gamma_r|f|^2,
\end{equation}
which is a third-order algebraic equation for mode energy $|a|^2$, where $\Delta \omega=\omega_0-\omega$ is the detuning frequency between the mode frequency  $\omega_0$ and laser frequency $\omega$, $\gamma = \gamma_r + \gamma_{nr}$ is a total loss rate of the mode comprising of nonradiative $\gamma_{nr}$ and radiative $\gamma_r$ contributions, $\alpha > 0$ is optical nonlinearity coefficient that defines BIC frequency shift under increasing temperature, and $|f|^2$ is the power of the incident wave. Three solutions of this equation correspond to bistability appearance in the system, whereas two solutions case define the borders of hysteresis area which could be obtained by setting derivative of the function $\zeta$ with respect to mode energy to zero $\zeta^{\prime}(|a|^2) = 0$. Inserting corresponding border values $|a_{\text{up, down}}|^2$ we acquire then switch power values:
\begin{equation}
    P_{\text{up, down}} =\frac{\zeta(|a_{\text{up, down}}|^2)}{\gamma_r} = \frac{2}{27}\dfrac{\gamma^3}{\alpha \gamma_r}\left( \dfrac{\Delta{\omega}}{\gamma}\left[\dfrac{\Delta{\omega}^2}{\gamma^2} + 9\right]\right) \pm \frac{2}{27} \dfrac{\gamma^3}{\alpha \gamma_r}\left(\dfrac{\Delta{\omega}^2}{\gamma^2} - 3\right)^{3/2},
    \label{eq_suppl: p_up_down}
\end{equation}
and the width of hysteresis area is then given by:
\begin{equation}
    W_{\text{h}} = P_{\text{up}} - P_{\text{down}} =\frac{4}{27} \dfrac{\gamma^3}{\alpha \gamma_r}\left(\dfrac{\Delta{\omega}^2}{\gamma^2} - 3\right)^{3/2}.
    \label{eq_suppl: hyst_width}
\end{equation}

\section{Nonlinear critical coupling}
One could see from Eq.~\ref{eq_suppl: hyst_width} that hysteresis width increases with increasing frequency detuning~$\Delta\omega$. However, in real physical systems this detuning could not be increased unlimetedly and is restricted by maximum available power $P_{max}$. Bistability regime is reached only when switch power $P_{up}$ is lower than this highest value. Therefore, as soon as power restriction condition is applied to our system, frequency detuning $\Delta\omega_{max}$ is no longer an independent parameter and should be derived from $P_{up} = P_{max}$ equation for every value of the losses rate. So, in this section we rigorously define the largest possible hysteresis width $W_{\text{h}}$ for the case of limited power in terms of radiative versus nonradiative losses ratio, which we refer to as \textit{nonlinear critical coupling}.

By implementing considerations given above and applying Eq.~\ref{eq_suppl: p_up_down},\ref{eq_suppl: hyst_width} we come to following expression:
\begin{equation}
    \frac{2}{27\alpha}\underbrace{\dfrac{\gamma^3}{\gamma_r}\left( \dfrac{\Delta{\omega}_{max}}{\gamma}\left[\dfrac{\Delta{\omega}_{max}^2}{\gamma^2} + 9\right]\right)}_{\text{term to minimize}} = \left(P_{max} - \frac{W_h}{2}\right),
\end{equation}
where the term on the left side should be minimized in terms of radiative losses $\gamma_r$ (we fix the nonradiative part $\gamma_{nr}$) for width maximization by taking the derivative and setting it to zero. Here, as mentioned above, $\Delta\omega_{max}$ is an implicit loss function determined by the maximum power condition. 

\begin{figure}[h!]
    \centering   \includegraphics[width=0.7\columnwidth]{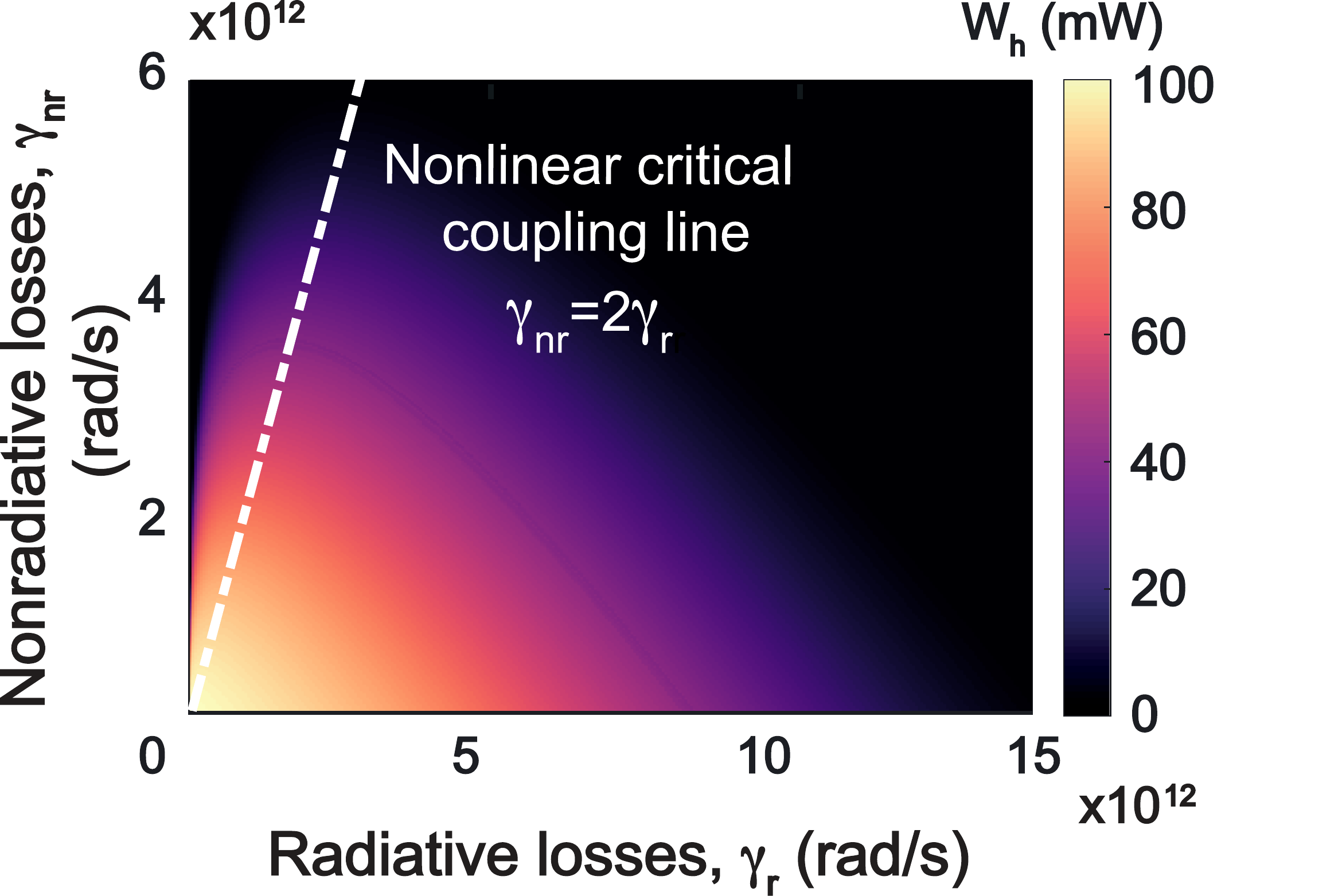}
    \caption{ Hysteresis width map depending on
radiative and nonradiative losses in real ranges of values under the condition $P_{up} = P_{max}=100$ mW. The
white line indicates the critical coupling condition for the loss $\gamma_{nr}=2\gamma_r$ that
maximizes the width of the hysteresis.}
     \label{fig5:suppl}
\end{figure}

Finally, after completing mathematical transformations this optimization gives us the following relation between radiave and nonradiative contributionsS for losses:
\begin{equation}
    \gamma_r = {\gamma_{nr}}/{2}.
\end{equation}
The resulting nonlinear critical coupling condition is depicted by the dotted white line on the map calculated from the hysteresis width Eq.~\ref{eq_suppl: hyst_width} (Fig.~\ref{fig5:suppl}), and indeed corresponds to width maximization depending on radiative and nonradiative loss at the frequency detuning $\Delta\omega_{max}(\gamma_r, \gamma_{nr})$.

In addition, one can note that $\gamma_r = \gamma_{nr}/2$ is the same ratio for losses as was discussed for the critical power minimization. That is a logical conclusion in a sense that for fixed maximum pump value the lowest power for bistability appearance provides the broadest range for detuning variations. Therefore, these two approaches give us exactly the same answer.
